# On the Influence of Delay Line Uncertainty in THz Time-Domain Spectroscopy

D. Jahn [1], S. Lippert[1], M. Bisi[2], L. Oberto[2], J. C. Balzer[1], M.Koch[1]

1) Physics Department, Phillips University Marburg, Renthof 5, 35032 Marburg, Germany

2) *Istituto Nazionale di Ricerca Metrologica, strada delle Cacce 91, 10135, Torino, Italy*

## Abstract

Terahertz time-domain spectroscopy (THz TDS) is a well-known tool for material analysis in the terahertz frequency band. One crucial system component in every time-domain spectrometer is the delay line which is necessary to accomplish the sampling of the electric field over time. Despite the fact that most of the uncertainty sources in TDS have been discussed, the delay line uncertainty has not been considered in detail. We model the impact of delay line uncertainty on the acquired THz TDS data. Interferometric measurements of the delay line precision and THz time-domain data are used to validate the theoretical model.

## Introduction

Terahertz time-domain spectroscopy (THz TDS) is a material analysis tool which is becoming more and more popular for basic research as well as for industrial applications. Extracting the optical constants of materials in the frequency band from 0.1 THz to 10 THz is one of the main goals of THz TDS. Contrary to the nearly perfect parameter extraction methods [1], [2] the measurement process itself will always suffer from measurement uncertainties, threatening the reliability of the measurements. Some contributions have already been discussed [3]–[6], while the impact of one crucial hardware component of a THz TDS system, the delay line, has not yet been investigated in detail. How precise should the delay line be if we want to have its impact on the overall measurement uncertainty neglected? For which inaccuracy will it be the limiting factor in THz TDS measurements?

In a classical TDS, the delay line introduces a variable time delay between the probing femtosecond (fs) laser pulse and the incident THz pulse, thus enabling the sampling of the latter. Any small deviation in the positioning of the delay line will therefore introduce a small uncertainty on the measured THz transient. In more detail, the measured quantity is proportional to the THz electric field $E(t)$ at the time $t$ of incidence of the fs laser pulse. The positioning error of the delay line will lead to an error in the time axis which directly translates to an error in the acquired electrical field $E(t)$. In the subsequent data analysis this will lead to errors in the spectrum $E(\omega)$ and thus in the material parameters, as e.g. the complex refractive index $n(\omega)$ [7]. The positioning noise has different origins such as sensor resolution, positioning reproducibility and mechanical limits of the delay line.

To answer the question whether the delay line uncertainty reduces the overall system performance we performed high precision delay line position measurements with an optical interferometer while acquiring THz time-domain traces. Experimentally we show that the amplitude noise of the THz field is affected by the delay line precision. Frequently, the THz TDS system performance is characterized by its maximal achievable Signal-to-Noise Ratio (SNR) in frequency domain which is defined as the maximal signal power $S$ divided by the average noise power $N$ [8]

$$SNR = \frac{S}{N} = S_{dB} - N_{dB}, \qquad (1)$$

where $S_{dB}$ and $N_{dB}$ are signal and noise power in a decibel scale.

It is worth noting that the time-domain signal and the spectrum $E(\omega)$ are related via Fourier transformation. Hence, we investigate the SNR change due to delay line noise, regarding the SNR as an indicator of system performance. We concentrate on random delay line noise and we do not deal with systematic time axis errors, *e.g.* periodic or constant deviation of the time axis, which can be minimized by a careful analysis of the measured time-domain data. For a periodic delay deviation this is most commonly done in frequency domain.

## Model

In the following we derive a basic model of the effect that a small random deviation in delay line position $\delta s$ has on the time axis and consequently on the measured electrical field $E(t)$. This will enable us to study the impact of delay line noise on the THz time-domain measurements. Figure 1) shows a typical THz TDS setup. A fs laser beam is split in two parts, one used for the photoconductive generation of THz radiation and the other for coherent detection of the THz pulse. The fs laser pulse, which is incident on a zinc-telluride crystal, probes the electrical field of the overlapped THz pulse via electro-optical sampling [9]. The delay line modifies the optical path length of the probing laser pulse and so its time of incidence on the detector. By moving the delay line it is possible to sample the THz electrical field in the time-domain.

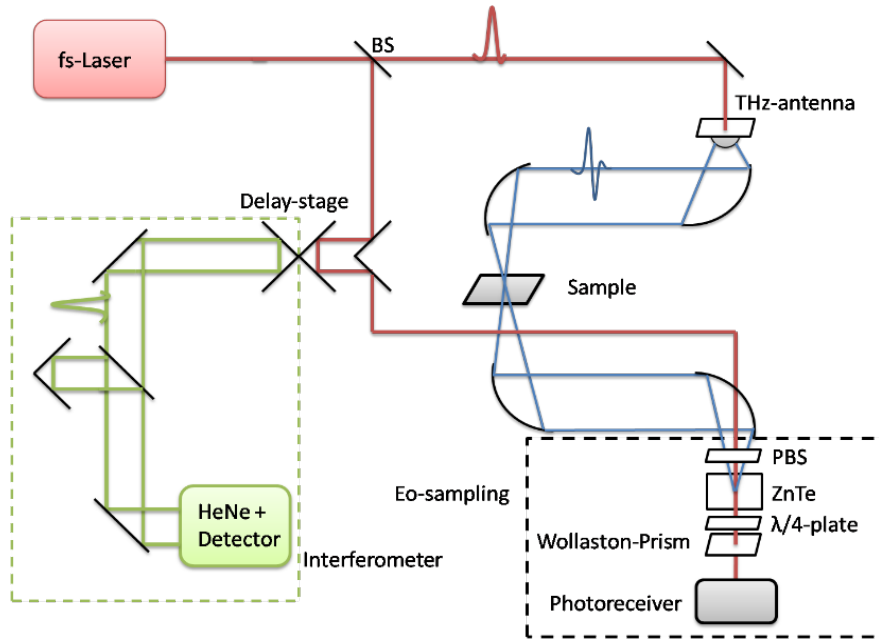

*Figure 1) Typical THz TDS setup. One part of the fs laser pulse is used to generate THz radiation by means of a photoconductive antenna while the second part is used for electro-optical sampling in a ZnTe crystal. The time delay between THz radiation and the probing laser pulse is altered by the delay line by changing the optical path length of the latter.*

Typically the delay line is equipped with a retroreflector. A little misplacement of the delay line by the amount $\delta s$ will delay the optical pulse by the additional time $\delta t$ as given by:

$$\delta t = \frac{2\delta s}{c}, \quad (2)$$

which results in an uncertainty in the time axis

$$\sigma_t = \frac{2}{c}\sigma_{position}. \quad (3)$$

Here $\delta s$ and $\delta t$ denote the position and time deviations, while $\sigma_{position}$ and $\sigma_t$ are the corresponding uncertainties. Since the delay line is used to sample the THz transient with the optical pulse, an error in its misplacement translates into an error in the measured amplitude of the electric field $E(t)$. If the fs laser pulse is delayed by an additional time $\delta t$, the electric field is measured as if it was at time $t + \delta t$, thus

$$E(t + \delta t) = E(t) + \delta t \frac{dE}{dt}. \quad (4)$$

From (4) we can derive the uncertainty in the electric field measurement from the law of propagation of uncertainties

$$\sigma_E(t) = \sigma_t |\frac{dE}{dt}(t)|. \quad (5)$$

Hence, the uncertainty $\sigma_E(t)$ of the electrical field depends on its slope. Since the spectrum $E(\omega)$ is the Fourier transform of the measured time-domain field $E(t)$, the uncertainty of the time-domain signal gets propagated through standard error propagation theory to the frequency domain [10]. To capture the impact of the delay line noise on the spectrum we use the maximal SNR as an indicator of system performance. Equation 5 shows that the noise amplitude scales linearly with the signal amplitude. As an important consequence it follows from equation (1) that the SNR will be independent of the signal power. A spectrometer's performance that is limited by delay line noise thus will never be improved by increasing the THz output power. As illustrated in figure 2) for the same delay line uncertainty $\sigma_{Position}$ the electric field uncertainty $\sigma_E$ increases with increasing pulse amplitude.

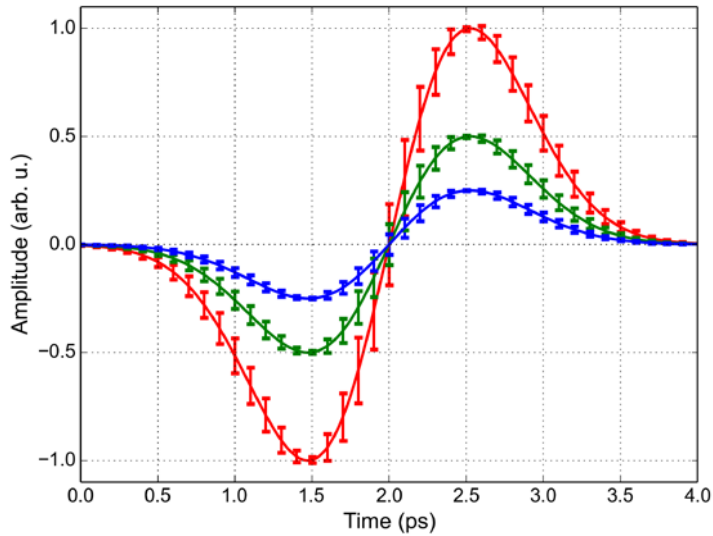

*Figure 2) For different signal amplitudes and equal delay line uncertainty the electric field uncertainty will scale linearly.*

## Monte Carlo Simulation

For a better understanding of the impact of delay line noise on the spectrum and the SNR, the analytical formulas are of little help. A better insight can be gained by means of Monte-Carlo (MC) simulations, which carry out the error propagation numerically. In this way the contribution of delay line noise can be studied separately, neglecting all other error contributions for sake of simplicity. An ideal pulse E(t) with unity amplitude and pulsewidth $t_E$ is used as initial pulse to perform the MC simulation. For each MC run white noise with standard deviation $\sigma_{position}$ is added to the position data. This leads to a noisy time-domain signal $E(t)$ and subsequently to a noisy spectrum $E(\omega)$. This step is repeated $N$ times until a statistically meaningful standard deviation of $E(t)$ and $E(\omega)$ can be calculated. The number of iterations $N$ was $10^5$ in our case. In time-domain, the simulations show that white noise on the time axis will lead to a time dependent standard deviation $\sigma_E(t) = |\frac{dE}{dt}|\sigma_t$ as expected from (5). For a given uncertainty in time, the electric field measurement is strongly affected in regions where the electric field is changing rapidly, while it is almost unchanged in the flat regions. This is illustrated schematically in figure 3), left hand side, for two different time positions. A time deviation $\delta t$ produces an error $\delta E$ in the electrical field measurement, which is dependent on the slope of the pulse. This MC result is in good agreement with the previous considerations. Further, the linear scaling of the standard deviation with the field amplitude is verified. These results show that our MC simulations reproduce the expected behavior in time-domain.

In frequency domain the impact of delay line noise is harder to grasp. The right-hand side of figure 3) show two MC runs for different delay line noise. We conclude that, as for every frequency independent noise signal, a constant noise floor is generated. The absolute value of the noise floor depends on the positioning uncertainty $\sigma_{Position}$. A high delay line uncertainty will lead to a high noise floor (a low SNR) and vice versa.

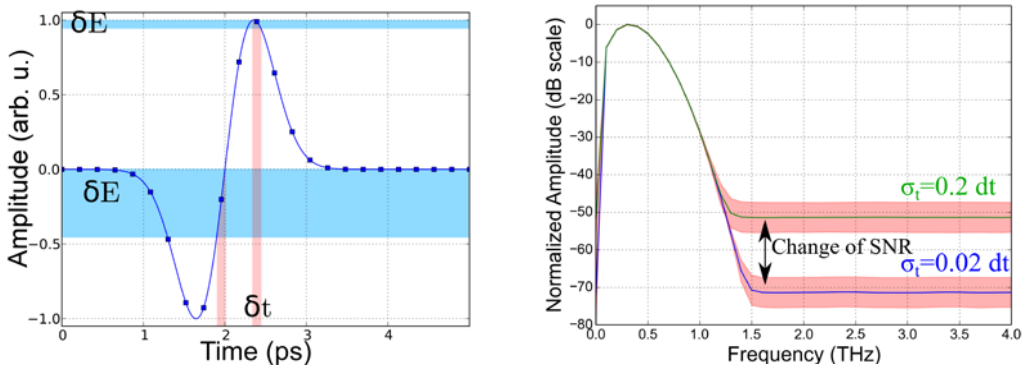

*Figure 3) Left: A deviation $\delta t$ in time translates into an error of the acquired time-domain signal $E(t)$. For a fixed time uncertainty, the uncertainty in the electrical field depends on the slope of the pulse. Right: Different time uncertainties $\sigma_t$ (20% and 2 % of the time step dt) generate a different noise floor in frequency domain. A given uncertainty in time will thus lead to a defined noise floor in frequency domain.*

From the MC simulations we conclude that in addition to the positioning uncertainty $\sigma_{Position}$, the sampling rate $1/\Delta t$ is influencing the value of the noise floor. A higher sampling rate decreases the noise floor. The results of the analysis are summarized in in figure 4). As can be seen, the maximal attainable SNR changes for a given time step. We observe a linear relationship between SNR and time step in the log-log plot. In other words, the SNR follows a power-law in time step $\Delta t$ with exponent one half and also scales linearly with the delay line uncertainty $\sigma_{Position}$. The SNRs calculated by the MC simulations are well fitted by the following equation:

$$SNR_{dB} = 20 log_{10}(C\sigma_t \sqrt{\Delta t}) . \qquad (6)$$

The contribution of $\sqrt{\Delta t}$ to the overall SNR can be explained with a statistical argument [11]. Each Fourier coefficient $E(\omega)$ is a sum of the noisy time-domain electrical field $E(t_i)$ and is given by $E(\omega) = \sum_i E(t_i) exp(-i\omega t_i)$. Decreasing the step size $\Delta t$ leads to an oversampling of $E(\omega)$. The number of summands increases significantly, without providing more information. Especially for the low frequencies the argument of the exponential function changes slowly, resulting in a kind of averaging of the noisy time-domain signal. From the central limit theorem it follows that in those cases $\sigma_E(\omega) \propto \frac{1}{\sqrt{N}} \propto \sqrt{\Delta t}$ which in turns is well confirmed by the good agreement of equation 6 with the MC data.

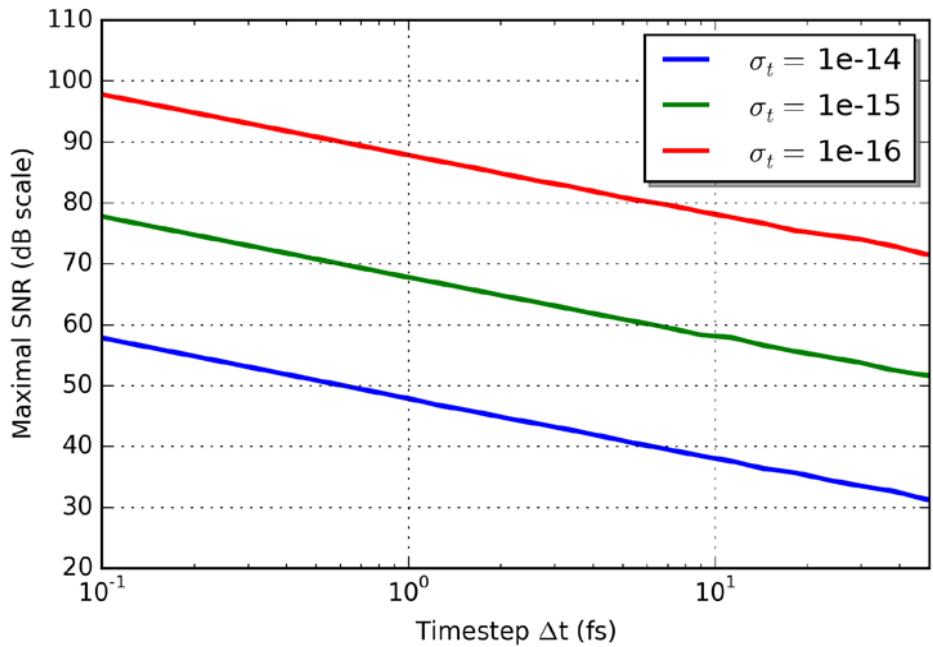

*Figure 4) Maximal SNR in dB for different time steps dt. Different curves depict different delay line uncertainty. The higher the sampling rate the better the maximal achievable SNR for each delay line noise.*

## Experimental results

In order to verify the numerical observations described above, we equipped the delay line of a THz TDS with an optical interferometer in order to perform precise measurement of the delay line position while acquiring the THz pulse. This enables us to compare the time axis obtained from the interferometric measurement with the time axis obtained from the delay line controller. For the forward and backward direction movement of the delay line, the Gaussian fit of the position deviation of the delay line controller value and the interferometer value yields to a standard deviation of $\sigma_{t,forward} = 1.3 fs$ and $\sigma_{t,backward} = 0.4 fs$ respectively. The used delay line controller limits the step width to 17 fs. The data has been recorded with this smallest possible time step. We observed a different positioning noise in the forward and in the backward direction. Figure 5) shows the distribution of the delay line noise after subtraction of all systematic errors.

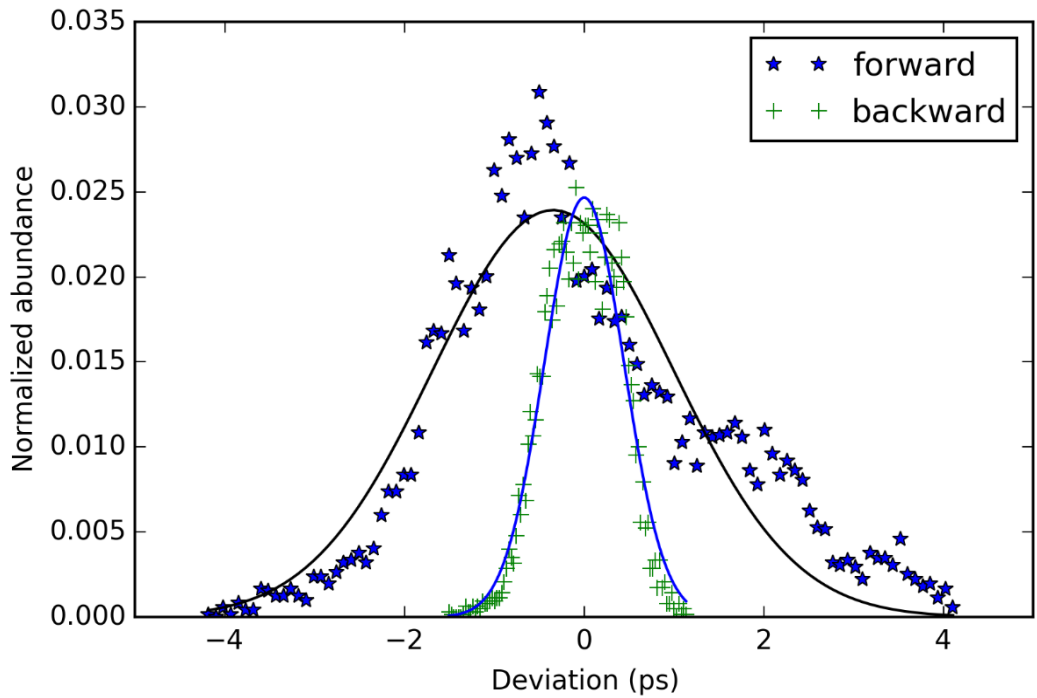

*Figure 5) Distribution of the deviation of the interferometer reading for the forward and backward motion of the delay line. The FWHM of the forward motion noise is approximately 3 times larger than that of the backward motion.*

The results of the interferometric measurements can be used to beforehand estimate the maximal SNR of our spectrometer, as limited by the delay line. The ideal pulse from the MC simulations is fitted to the measured pulse shape. Our MC simulations predict that the spectrometer will not exceed a SNR of 63 dB in forward direction and a SNR of 73 dB in backward direction. In order to confirm this prediction experimentally, we have recorded the THz time-domain signals in both directions. The interferometric measurements suggest, that we will have a worse movement of the delay line in forward than in backward direction. We thus expect a change of SNR between the two directions. The corresponding spectra are shown in figure 6).

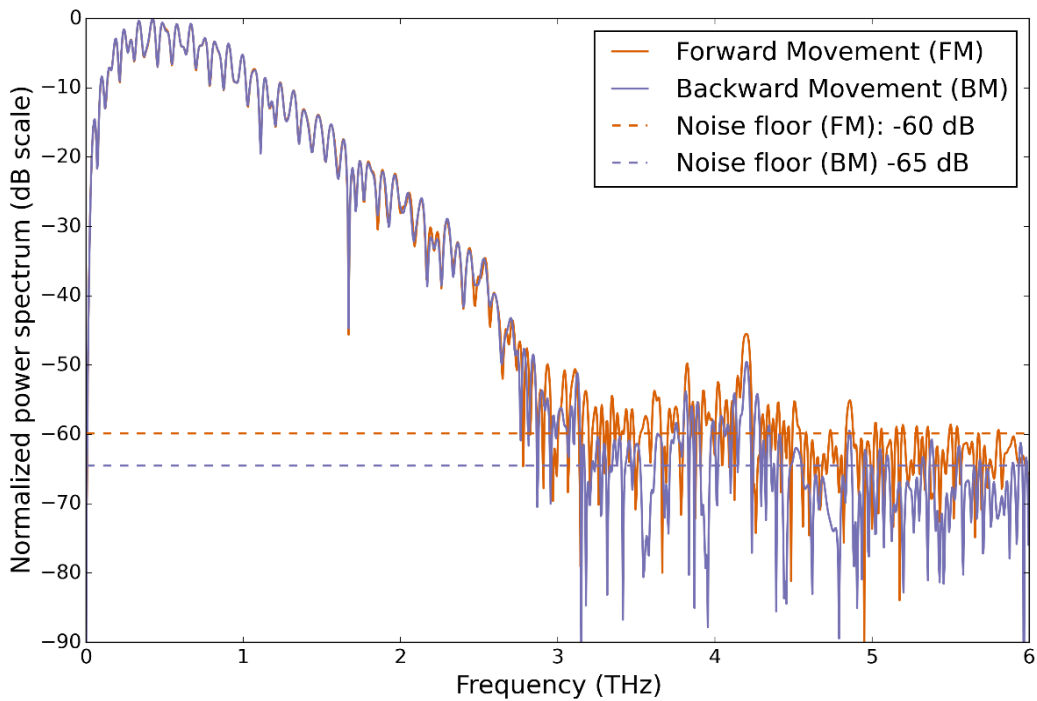

*Figure 6) Normalized spectra for forward and backward movement. In either direction (forward, backward) ten time-domain signals are averaged. The noise floor for the two directions differs by 5 dB. This indicates an influence of the delay line movement on the measured THz-field.*

We observe a 5 dB difference in the noise floor for the movement in forward and backward direction. In case of backward movement, where theoretically a SNR of 73 dB should have been achieved, other sources of error seem to limit the SNR of the system. An even more precise movement of the delay line will not increase the systems performance further. In contrast to this finding, for the forward movement the delay line noise is obviously larger than that of other error sources, since an improvement of SNR can be achieved by switching to the more precise time axis.

## Conclusion

In conclusion we have developed a model that describes how the uncertainty in delay line positioning propagates to the acquired THz time-domain signal and further to its spectrum. According to the Nyquist theorem, the sampling time interval $\Delta t$ gives the maximal frequency $f_{max} = \frac{1}{2\Delta t}$ and it might be obvious to choose the sampling time such that $f_{max}$ is slightly above the expected bandwidth of the spectrometer. However, we have shown that the SNR can be considerably reduced by oversampling of the THz pulse. Hence, in a single measurement it is advisable to use a reasonably low time interval. The absolute value of the acceptable delay line noise depends on several contributions. First the smallest usable sampling time, or similarly the number of averages of full time-domain traces that can be taken. Second, the bandwidth of the system, since a more precise delay line matters at high frequencies, where typically the SNR of the system is quite low. A pulse with high frequency components is necessarily short in time-domain, hence the slope of the pulse is large. Given that the delay line noise is proportional to the slope in time-domain, the impact of delay line noise is more severe for higher frequencies.

In summary we have shown that in a single measurement a high SNR can be obtained by using a high precision delay line. In principle also delay lines with a lower precision can yield to an overall acceptable SNR if either averaging or oversampling of pulse signals can be performed. It is worth noting that oversampling, or averages, must be traded off in order to obtain an acceptable measurement speed.

## Acknowledgement

D.J., S.L., M.B. and L.O. acknowledge financial support by the EMRP NEW07, that is jointly funded by the EMRP participating countries within EURAMET and the European Union. D. J. acknowledges financial support by Ev. Studienwerk e.V.